\documentclass[fleqn]{SCPMA}

\usepackage{bm}
\usepackage{algorithm}
\usepackage{multirow}
\usepackage{float}
\usepackage[numbers,sort&compress]{natbib}
\usepackage{orcidlink}

\begin{document}
\renewcommand{\thefootnote}{\fnsymbol{footnote}}
\raggedbottom
\sloppy

\ensubject{subject}

\ArticleType{Article}
\Year{2026}
\Month{January}
\Vol{66}
\No{1}
\DOI{??}
\ArtNo{000000}
\ReceiveDate{January 1, 2026}
\AcceptDate{June 1, 2026}

\title{MeV Gamma-Ray Lines from Radioactive Nuclei in Magnetar Giant Flares}{MeV Gamma-Ray Lines from Radioactive Nuclei in Magnetar Giant Flares}

\author[1]{Wu-Zimo Qiumu \orcidlink{0009-0003-7350-7128}}{}
\author[1]{Meng-Hua Chen \orcidlink{0000-0001-8406-8683}
\thanks{Corresponding author (M.-H. Chen, email: mhc@gxu.edu.cn)}}{} 
\author[1]{Qiu-Hong Chen \orcidlink{0009-0006-8625-5283}}{}
\author[1]{Fei Xie \orcidlink{0000-0002-0105-5826}}{}
\author[1]{\protect\\ Hou-Jun L\"u \orcidlink{0000-0001-6396-9386}}{}
\author[1]{Xiang-Gao Wang \orcidlink{0000-0001-8411-8011}}{}
\author[1]{En-Wei Liang \orcidlink{0000-0002-7044-733X}}{}

\AuthorMark{W.-Z. Qiumu}

\AuthorCitation{W.-Z. Qiumu, M.-H. Chen, Q.-H. Chen, F. Xie, H.-J. L\"u, X.-G. Wang, and E.-W. Liang}

\address[1]{Guangxi Key Laboratory for Relativistic Astrophysics, School of Physical Science and Technology, Guangxi University, Nanning 530004, China}

\abstract{
The rapid neutron-capture process ($r$-process) is widely regarded as the dominant mechanism responsible for the synthesis of heavy elements in the universe, yet its astrophysical sites remain an open question. Recent studies suggest that the high-entropy, rapidly expanding baryonic material ejected by magnetar giant flares may provide favorable conditions for $r$-process nucleosynthesis, while the late-time gamma-ray emission observed from the magnetar SGR~1806-20 offers direct observational support for this scenario. In this work, we perform nuclear reaction network simulations to investigate the nucleosynthesis yields of magnetar giant flares and to characterize the associated nuclear gamma-ray line emission arising from the radioactive decay of heavy nuclei. The nuclei synthesized in magnetar giant flares are found to be mainly distributed near the first and second $r$-process abundance peaks. Owing to this nuclide composition, the gamma-ray opacity is found to be strongly energy-dependent with the opacity in the keV band exceeding that in the MeV band by approximately three orders of magnitude. The nuclear gamma-ray emission is dominated by MeV photons at early times ($t \lesssim 10^3$~s) and gradually extends toward the sub-MeV and keV bands as time progresses, thereby offering a diagnostic of heavy element enrichment in the ejecta. The gamma-ray spectrum exhibits a peak near $\sim 1$~MeV with major contributions from $^{88}$Kr and $^{92}$Sr, whose radioactive decays produce several bright gamma-ray lines with fluxes exceeding $\sim10^{-8}$~erg~cm$^{-2}$~s$^{-1}$, making them the most promising lines for detection by MeV gamma-ray detectors. Because magnetar giant flares occur in the Galaxy at a rate roughly three orders of magnitude higher than neutron star mergers and their nuclear gamma-ray lines are accessible to current MeV instruments, they offer new and valuable science opportunities for MeV gamma-ray astronomy.
}



\maketitle

\begin{multicols}{2}

\section{Introduction}

The rapid neutron-capture process ($r$-process) has long been recognized as the dominant mechanism responsible for the synthesis of heavy elements beyond iron in the universe~\citep{1957RvMP...29..547B}. However, the astrophysical sites where the $r$-process occurs remain an open question~\citep{2021RvMP...93a5002C}. The $r$-process requires extreme nuclear reaction conditions characterized by high temperatures, high densities, and neutron-rich environments. Mergers of compact star, such as neutron star - neutron star mergers~\citep{1974ApJ...192L.145L,1982ApL....22..143S}, neutron star - black hole mergers~\citep{2021LRR....24....5K}, and even neutron star - white dwarf mergers~\citep{2016MNRAS.461.1154M,2025ApJ...988L..46L}, as well as core-collapse supernovae~\citep{1994ApJ...433..229W,1996ApJ...471..331Q}, are considered among the most plausible $r$-process sites based on numerical simulations. In particular, the neutron-rich ejecta from neutron star mergers are considered the most promising sites for the production of heavy $r$-process nuclei~\citep{2012MNRAS.426.1940K,2013PhRvD..87b4001H}. The radioactive decay of freshly synthesized $r$-process nuclei generates various decay products, including $\alpha$-particles, $\beta$-particles, gamma-rays, and neutrinos~\citep{2010MNRAS.406.2650M,2024MNRAS.527.5540C}. Charged particles such as $\alpha$ and $\beta$-particles efficiently thermalize their energy within the ejecta, powering bright thermal emission known as a kilonova~\citep{1998ApJ...507L..59L,2010MNRAS.406.2650M}. In contrast, gamma-rays and neutrinos can be partially scattered or absorbed by the ejecta, while a fraction of them may escape directly from the ejecta, giving rise to radioactive gamma-ray emission and neutrino emission~\citep{2016MNRAS.459...35H,2019ApJ...872...19L,2020ApJ...889..168K,2020ApJ...903L...3W,2021ApJ...919...59C,2022ApJ...932L...7C,2023MNRAS.520.2806C}. Notably, radioactive gamma-ray emission, which originates from nuclear level transitions, represents a distinctive probe of $r$-process nucleosynthesis and is a key topic in MeV gamma-ray astronomy~\citep{2022PrPNP.12703983D}. Radioactively powered kilonova emission was observed in the first gravitational wave detected neutron star merger event, GW170817~\citep{2017ApJ...848L..12A}, providing strong evidence that neutron star mergers are major production sites of heavy elements in the universe~\citep{2017Natur.551...80K,2024MNRAS.529.1154C}. However, the detection of characteristic gamma-ray line emission arising from nuclear transitions in neutron star merger ejecta remains a significant challenge for current MeV gamma-ray instruments~\citep{2021ApJ...919...59C,2022ApJ...932L...7C,2024ApJ...971..143C}.

Beyond neutron star mergers, magnetar giant flares (GFs) have recently been proposed as another potential site for $r$-process nucleosynthesis~\citep{2024MNRAS.528.5323C}. Hydrodynamical simulations indicate that magnetar GFs can eject $\sim 10^{-8}-10^{-6}~M_\odot$ of high-entropy and rapidly expanding baryonic material, providing favorable conditions for the synthesis of light $r$-process elements~\citep{2025ApJ...985..234P}. 
This ejecta mass range is also supported by observations and modeling of synchrotron radio afterglows~\citep{2005Natur.434.1112C,2005Natur.434.1104G,2005ApJ...634L..93T,2005ApJ...634L..89G,2006ApJ...638..391G}.
Recent work~\citep{2025ApJ...984L..29P} further investigated the gamma-ray emission driven by the radioactive decay of heavy elements in magnetar GFs and found that the predicted late-time gamma-ray signatures are broadly consistent with the delayed gamma-ray features observed from the magnetar SGR~1806-20~\citep{2005ApJ...624L.105M,2005Natur.434.1098H,2007ApJ...661..458B,2007AstL...33....1F}.
These results suggest that magnetar GFs could make a non-negligible contribution to $r$-process nucleosynthesis. Therefore, it is worthwhile to further investigate their nucleosynthetic yields and the associated nuclear gamma-ray line emission.

Compared with neutron star mergers, the mass of material ejected by magnetar GFs is smaller by $\sim4-6$ orders of magnitude and their contribution to the production of heavy elements is expected to be relatively modest. A recent study estimated the event rate of such flares and the nucleosynthesis yield per event, finding that magnetar GFs may contribute $\sim1-10\%$ of the total Galactic $r$-process budget~\citep{2025ApJ...985..234P}. However, the occurrence rate of magnetar GFs in the Milky Way is higher than that of neutron star mergers by a similar factor of $\sim4-6$ orders of magnitude. As a result, magnetar GFs are expected to make a significant contribution to the Galactic MeV gamma-ray background, providing a promising source for characteristic gamma-ray line emission. Previous work~\citep{2025ApJ...984L..29P} estimated the spectral features of the nuclear gamma-ray emission from magnetar GFs by assuming a constant gamma-ray opacity without accounting for the detailed nuclear composition of the ejecta. In reality, the gamma-ray opacity is strongly energy-dependent, which can substantially influence the spectral evolution of the gamma-ray emission~\citep{2021ApJ...919...59C,2022ApJ...932L...7C}. In this work, we use $r$-process simulations to study the nucleosynthetic yields of magnetar GFs and adopt energy-dependent gamma-ray opacities to investigate the properties of the resulting gamma-ray line emission and to identify the nuclei that contribute most significantly to the prominent gamma-ray lines.

The paper is organized as follows. In Section~\ref{Methods}, we describe the $r$-process nucleosynthesis calculations and the procedure used to model gamma-ray emission resulting from the radioactive decay of $r$-process nuclei. Section~\ref{Results} presents the predicted gamma-ray emission from magnetar GFs. Finally, Section~\ref{Summary} provides a summary of our findings and a discussion of their implications.

\section{Methods}
\label{Methods}

\subsection{Nucleosynthesis}

To simulate the evolution of nuclear abundances, we perform $r$-process nucleosynthesis calculations using the nuclear reaction network code SkyNet~\citep{2017ApJS..233...18L}. This code is a general-purpose nuclear reaction network that has been widely used in studies of diverse astrophysical scenarios, including $r$-process nucleosynthesis in magetar GFs~\citep{2025ApJ...985..234P}. The network includes $7843$ nuclide species and $\sim1.4\times10^5$ nuclear reactions. Nuclear reaction rates are taken from the JINA REACLIB database~\citep{2010ApJS..189..240C}. Nuclear masse values and radioactive decay data are obtained from AME2020~\citep{2021ChPhC..45c0003W} and NUBASE2020~\citep{2021ChPhC..45c0001K}, respectively.  For nuclide species lacking experimental data, we use theoretical data derived from the Finite-Range Droplet Model~\citep{2016ADNDT.109....1M}. Neutron capture rates are calculated using the Hauser-Feshbach statistical code TALYS~\citep{2008A&A...487..767G}. 

Following the $r$-process nucleosynthesis simulations for magnetar GFs~\citep{2025ApJ...985..234P}, we adopt the density profile
\begin{equation}
\rho (t) = 
\begin{cases}
\rho_{\rm D} \left( \frac{t}{t_{\rm D}} \right)^{-1}, & t_0 \leq t \leq t_{\rm D}, \\
\rho_{\rm D} \left( \frac{t}{t_{\rm D}} \right)^{-3}, & t > t_{\rm D},
\end{cases}
\end{equation}
where $t_{\rm D}=2R_{\rm NS}/v$ is the time for the layer to expand by a distance comparable to the neutron star diameter. The density normalization for each velocity layer is calcuated as
\begin{equation}
    \rho_{\rm D} = \frac{3}{4\pi} \frac{M_{\rm ej}}{ (\bar{v} t_{\rm D})^{3}} \left(\frac{v}{\bar{v}}\right)^{-6},
\end{equation}
where $M_{\rm ej}$ is the mass of ejected materials, $v$ is the expansion velocity, and $\bar{v}$ is the characteristic minimum velocity.

The astrophysical inputs used in our $r$-process nucleosynthesis are adopted from the detailed numerical simulations~\citep{2024MNRAS.528.5323C}. We set the initial electron fraction to be consistent with that of an iron nucleus, i.e., $Y_{\rm e}=26/(26+30)\approx0.46$. The specific entropy $s$ of each layer is computed under the assumption of nuclear statistical equilibrium using the Helmholtz equation of state. The initial temperature of the layer is then determined from its density and specific entropy. According to the numerical simulations~\citep{2024MNRAS.528.5323C}, the entropy of the ejecta from magnetar GFs can reach $s \sim$ several hundred $k_{\rm B}$ per baryon, with expansion velocity up to $\sim 0.3c$. Such high entropy and rapid expansion timescale provide condition favorable for the production of heavy elements via $r$-process nucleosynthesis.

\subsection{Gamma-Ray Emission}

The gamma-ray energy generation rate produced by the radioactive decay of heavy nuclei can be obtained by~\citep{2021ApJ...919...59C}:
\begin{equation}
    \dot{E}_{\gamma} (t) = N_{\rm A} \sum_{i} Y_{i}(t) \sum_{j} \lambda_{ij} b_{ij} \sum_{k} \varepsilon_{ijk} I_{ijk},
\end{equation}
where $N_{\rm A}$ is Avogadro's number, $Y_i(t)$ is the abundance of the $i$th nuclide, $\lambda_{ij}$ is the decay constant for the $j$th decay mode of the $i$th nuclide, $b_{ij}$ is the branching ratio, $\varepsilon_{ijk}$ is the energy of the $k$th photon produced in the $j$th decay mode, and $I_{ijk}$ is the corresponding intensity. In our calculations, gamma-ray radiation data for unstable nuclei are taken from the latest version of the nuclear data library ENDF/B-VIII.1\footnote{https://www.nndc.bnl.gov/endf/}.

To calculate the gamma-ray emission from the ejcted material, we account for absorption and scattering processes within the ejecta. In particular, we include four interaction channels between gamma-ray photons and matter: photoelectric absorption, Compton scattering, pair production, and Rayleigh scattering. The total opacity depends on the nuclide composition of the ejecta. In our model, it is calculated as the mass-fraction-weighted average~\citep{2022ApJ...932L...7C}:
\begin{equation}
    \kappa (\varepsilon) = \sum_i X_i \kappa_i (\varepsilon),
\end{equation}
where $\kappa_i (\varepsilon)$ is the opacity of the $i$th nuclide and $X_i$ is its
mass fraction. Opacity values for elements from hydrogen ($Z=1$) to fermium ($Z=100$) are taken from the XCOM database\footnote{https://www.nist.gov/pml/xcom-photon-cross-sections-database}. We note that the database provides opacity values only at discrete photon energies rather than as a continuous function. In order to study the interaction of gamma-ray photons across the full energy range with the dense ejecta, we represent the opacity using the following polynomial form:
\begin{equation}
    \log \kappa(E_{\gamma}) = C_0  + C_1 \log E_{\gamma} + C_2 (\log E_{\gamma})^2 + C_3 (\log E_{\gamma})^3,
\label{function}
\end{equation}
where the coefficients $C_0 - C_3$ are obtained by fitting the tabulated XCOM opacity values.

Considering that the ejecta produced during magnetar GFs typically cover a solid angle smaller than $4\pi$, the density profile can be expressed as
\begin{equation}
    \rho (r,t) = \frac{3}{4\pi f_{\Omega}} \frac{M_{\rm ej}}{ (\bar{v} t)^{3}} \left(\frac{v}{\bar{v}}\right)^{-6} = \frac{3M_{\rm ej}}{4\pi f_{\Omega}} \frac{(\bar{v}t)^3}{r^6},
\end{equation}
where $f_{\Omega}\equiv\Delta\Omega/4\pi$ is the outflow covering fraction, and $\Delta\Omega\le4\pi$ is the solid-angle subtended by the ejecta. The optical depth of the ejecta is given by
\begin{equation}
    \tau_{\gamma} (E_{\gamma},r,t) = \int_r^{\infty} \rho(r', t) \kappa(E_{\gamma}) dr' = \frac{3M_{\rm ej}\kappa(E_{\gamma})}{20\pi f_{\Omega}} \frac{(\bar{v}t)^3}{r^5},
\end{equation}
so that the gamma-ray photosphere ($\tau_{\gamma}=1$) is located at
\begin{equation}
    r_{\rm ph}(E_{\gamma},t) = \left( \frac{3M_{\rm ej}\bar{v}^3t^3\kappa(E_{\gamma})}{20\pi f_{\Omega}} \right)^{1/5}.
\end{equation}
When the ejecta become optically thin, such that the photosphere velocity is comparable to the characteristic minimum velocity ($v_{\rm ph}\approx\bar{v}$), the corresponding timescale is given by
\begin{equation}
    t_{\rm peak} \approx \left(\frac{3M_{\rm ej}\kappa(E_{\gamma})}{20\pi f_{\Omega}\bar{v}^2}\right)^{1/2} \approx 10^3~{\rm s} \left(\frac{M_{\rm ej}/f_{\Omega}}{10^{-6} M_{\odot}}\right)^{1/2} \left(\frac{\kappa(E_{\gamma})}{0.1~{\rm cm^2~g^{-1}}}\right)^{1/2} \left(\frac{\bar{v}}{0.1c}\right)^{-1}.
\end{equation}

The total gamma-ray luminosity powered by the radioactive decay of heavy elements in the magnetar GFs can be obtained by
\begin{equation}
    L_{\gamma} (t) = M_{\rm ej} N_A \sum_{i} Y_{i}(t) \sum_{j} \lambda_{ij} b_{ij} \sum_{k} \varepsilon_{ijk} I_{ijk} e^{-\tau_{\gamma}}.
\end{equation}
The corresponding photon flux can be written as
\begin{equation}
    F_{\gamma} (t) = \frac{L_{\gamma} (t)}{4\pi d^2},
\end{equation}
where $d$ is the distance to the source. We adopt $d=8.7$~kpc which is consistent with the distance to the magnetar SGR~1806-20~\citep{2008MNRAS.386L..23B}.

Considering that the ejecta from magnetar GFs expand at a high velocity, gamma-ray emission lines are expected to be Doppler broadened. Following the method adopted in earlier studies~\citep{2016MNRAS.459...35H}, we model each gamma-ray line by convolving it with a Gaussian profile
\begin{equation}
    \phi (E) = \frac{1}{\sigma\sqrt{2\pi}}\exp{\left(-\frac{(E-E_{\gamma})^2}{2\sigma^2}\right)},
\end{equation}
where $\sigma = 2\sqrt{\ln 2} v/c$ is the Doppler broadening  width.

\begin{figure*}[htbp]
    \centering
    \includegraphics[width=0.65\textwidth]{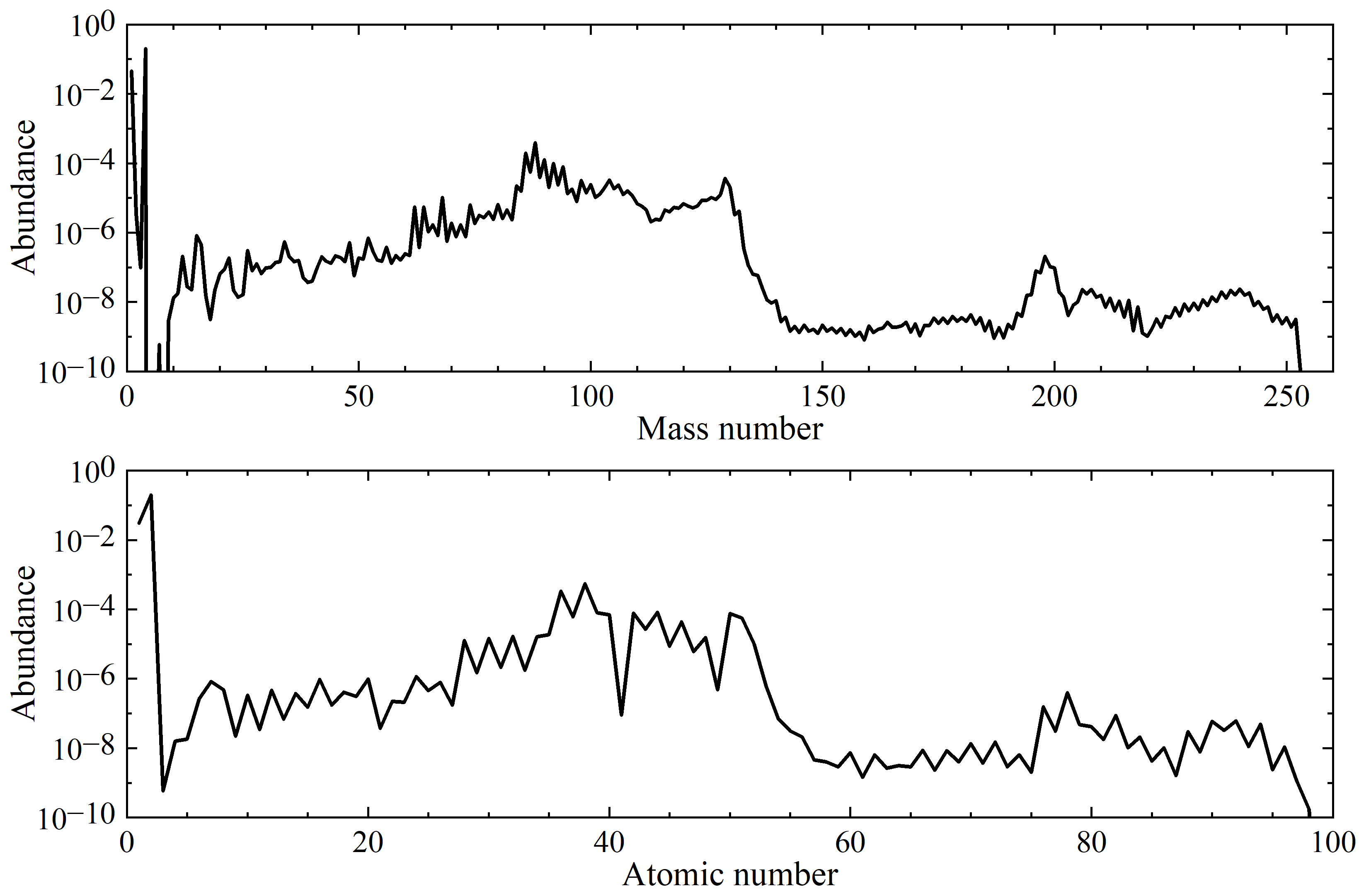}
    \caption{Abundance patterns of nuclei synthesized in a magnetar GF as functions of mass number (top panel) and atomic number (bottom panel) at $t=10^3$~s.}
    \label{abundance}
\end{figure*}

\section{Results}
\label{Results}

\subsection{Nucleosynthesis Yields of Magnetar GFs}

Figure~\ref{abundance} shows the abundance patterns of nuclei synthesized in a magnetar GF from $r$-process nucleosynthesis simulations. It can be seen that the synthsized nuclei are mainly distributed in the range of atomic numbers $25 \leq Z \leq 55$, corresponding to mass numbers $60 \leq A \leq 130$. Distinct peaks appear around $A \sim 80 - 90$ and $A \sim1 30$, consistent with the first and second $r$-process abundance peaks in the solar system, respectively. A weak peak is also seen around $A\sim195$, corresponding to the third $r$-process peak, although its abundance is about two orders of magnitude lower than those of the first and second $r$-process peaks. Our nucleosynthesis simulation results are broadly consistent with the $r$-process yields from magnetar GFs~\citep{2025ApJ...985..234P}, except for nuclei with mass numbers $A \gtrsim 200$. This discrepancy is likely attributable to our use of updated nuclear physics inputs and has little effect on the nuclear gamma-ray emission as the abundances of these heavy nuclei are extremely low.

\begin{figure}[H]
    \centering
    \includegraphics[width=1.0\linewidth]{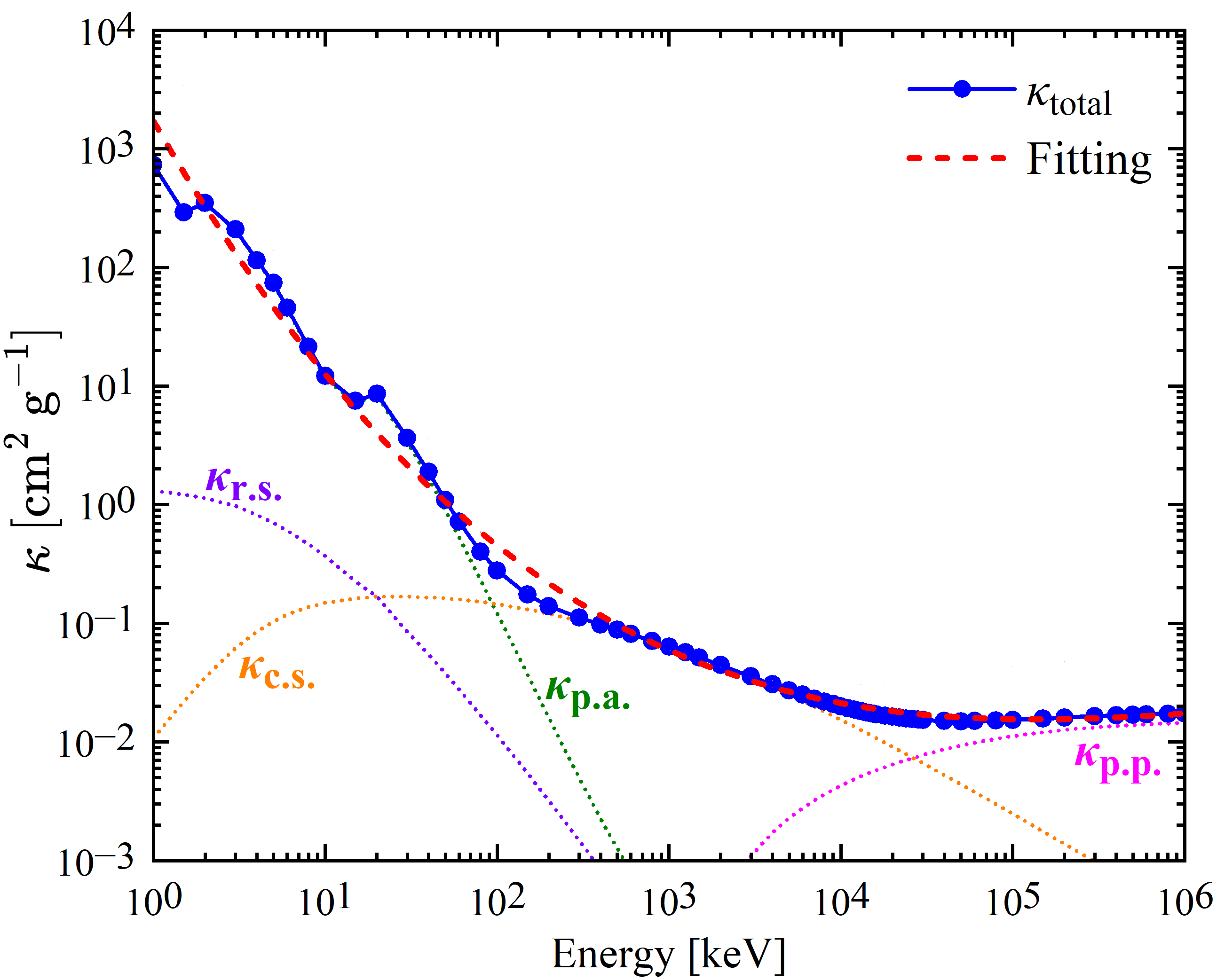}
    \caption{Gamma-ray opacity of the $r$-process material ejected in magnetar GFs. The total opacity ($\kappa_{\rm total}$) includes contributions from four interaction processes between gamma-rays and matter: photoelectric absorption ($\kappa_{\rm p.a.}$), Compton scattering ($\kappa_{\rm c.s.}$), pair production ($\kappa_{\rm p.p.}$), and Rayleigh scattering ($\kappa_{\rm r.s.}$). The red dashed line shows the opacity obtained by fitting the XCOM data with the polynomial function (Eq.~\ref{function}).}
    \label{kappa}
\end{figure}

\subsection{Gamma-Ray Opacity}

Based on the nuclide composition obtained from our $r$-process simulations, we calculate the total gamma-ray opacity of the ejecta from magnetar GFs, as shown in Figure~\ref{kappa}. We consider four parimary interaction processes between gamma-rays and matter: photoelectric absorption, Compton scattering, pair production, and Rayleigh scattering. It can be observed that the gamma-ray opacity increases significantly toward lower photon energies. For photon energies $E_{\gamma} \lesssim 100$~keV, the interaction of gamma-rays with matter is dominated by photoelectric absorption, which exceeds the contributions from Rayleigh and Compton scattering by several orders of magnitude. For photon energies between $100$~keV and $10$~MeV, the opacity is dominated by Compton scattering, while pair production in the nuclear field becomes significant above $10$~MeV. Figure~\ref{kappa} also shows the polynomial fit to the discrete opacity values obtained from the XCOM database, which reproduces the tabulated data well. The corresponding polynomial coefficients are listed in Table~\ref{para}. The gamma-ray opacity in magnetar GFs is broadly similar to but slightly lower than that in neutron star mergers~\citep{2016ApJ...829..110B,2021ApJ...919...59C,2022ApJ...932L...7C}.

\begin{table}[H]
    \centering
    \caption{Fitted polynomial coefficients for the total gamma-ray opacity.}
    \begin{tabular}{ccc}
    \hline\hline
    Coefficient   &   Value   &   Error   \\
    \hline
    $C_0$   &   $ 3.232$   &   $0.051$   \\
    $C_1$   &   $-2.518$   &   $0.055$   \\
    $C_2$   &   $ 0.407$   &   $0.016$   \\
    $C_3$   &   $-0.021$   &   $0.001$   \\
    \hline
    \end{tabular}
    \label{para}
\end{table}

\begin{figure}[H]
    \centering
    \includegraphics[width=1.0\linewidth]{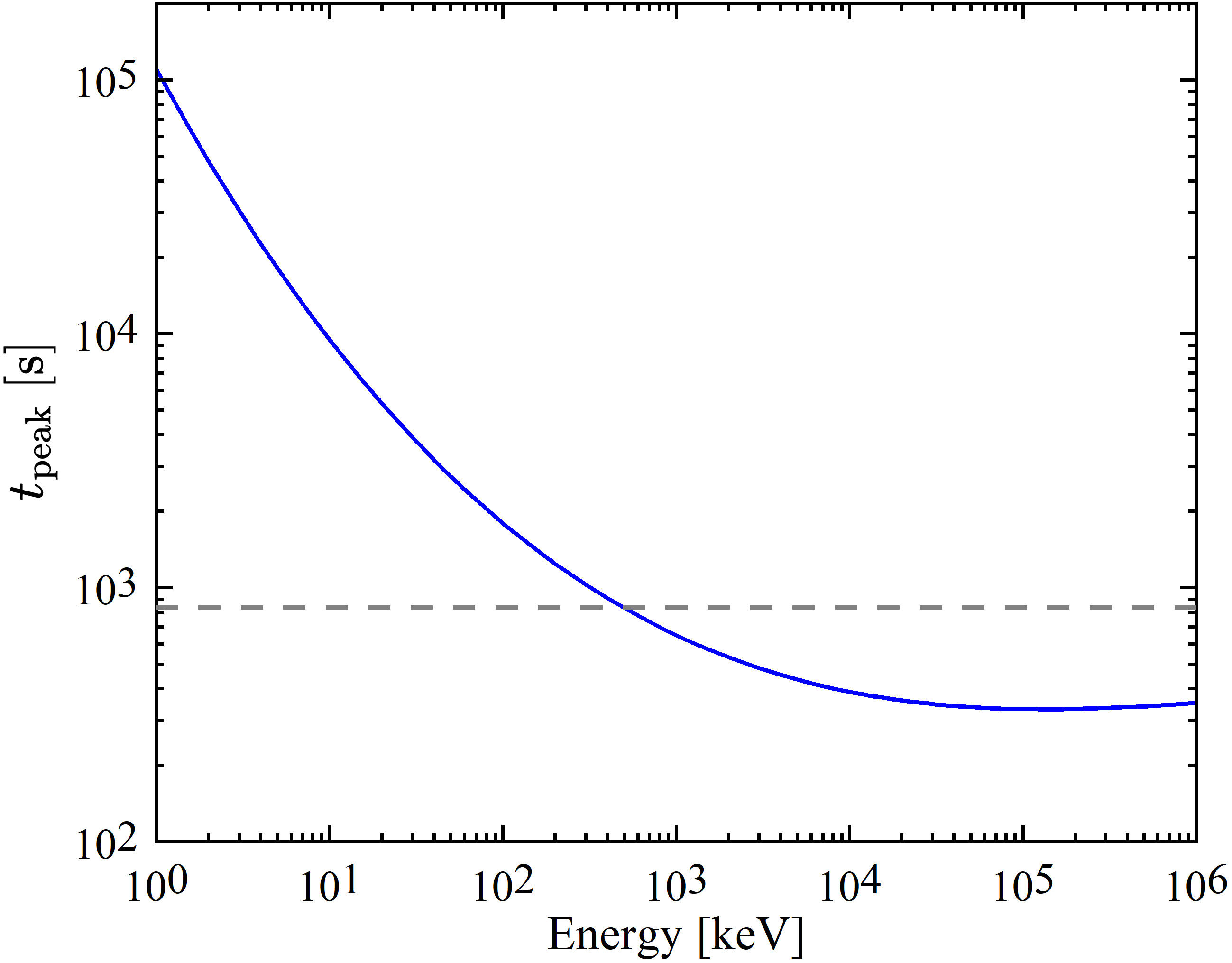}
    \caption{Peak timescale of gamma-ray emission as a function of photon energy. The dashed line denotes the peak timescale calculated by assuming a constant gamma-ray opacity of $\kappa = 0.1~{\rm cm^{2}~g^{-1}}$. The ejecta properties are adopted to be consistent with those inferred from the gamma-ray emission of SGR~1806-20.}
    \label{tpeak}
\end{figure}

The opacity of gamma-rays strongly influences their radiation transport across different photon energies. We therefore investigate the peak emission timescale of gamma-ray signal as a function of photon energy, as shown in Figure~\ref{tpeak}. It is found that the peak timescale is strongly sensitive to the gamma-ray opacity. Due to the relatively small interaction cross sections of high-energy gamma-rays (e.g., $E_{\gamma} \gtrsim 1$~MeV) with matter, these photons can escape from the ejecta at earlier times, leading to a peak timescale of $t_{\rm peak} \sim 300$~s. In contrast, for low-energy gamma-rays ($E_{\gamma} \lesssim 1$~MeV), the gamma-ray opacity increases sharply toward lower photon energies, causing their peak timescale to be substantially delayed, typically reaching $t_{\rm peak} \sim 10^{3}-10^{4}$~s. Consequently, gamma-rays in the MeV energy range are expected to dominate at early times, while the observed spectrum gradually extends toward sub-MeV and keV ranges as time progresses.

It is worth noting that previous work~\citep{2025ApJ...984L..29P} assumed a constant gamma-ray opacity of $\kappa = 0.1~{\rm cm^{2}~g^{-1}}$ in their radiation transport calculations. For comparison, Figure~\ref{tpeak} also shows the peak timescale derived using this constant opacity (shown as the dashed line). It is found that assuming a constant opacity causes photons at all energies to exhibit the same peak time, thereby eliminating the intrinsic energy-dependent delay produced by realistic opacity. In the photon energy range of $E_{\gamma} \approx 0.1-1$~MeV, the peak timescales computed with $\kappa = 0.1~{\rm cm^{2}~g^{-1}}$ are broadly consistent with those obtained using the realistic opacity. However, at lower energies ($E_{\gamma} \lesssim 0.1$~MeV), this assumption introduces substantial discrepancies. The peak timescale predicted using the realistic opacity is delayed by approximately one to two orders of magnitude relative to that computed with the constant opacity. Therefore, adopting an energy-dependent opacity is essential for accurately modeling gamma-ray radiation transport in the ejecta composed of heavy nuclei.

\subsection{Gamma-Ray Emission}

\begin{figure*}[t!]
    \centering
    \includegraphics[width=0.7\textwidth]{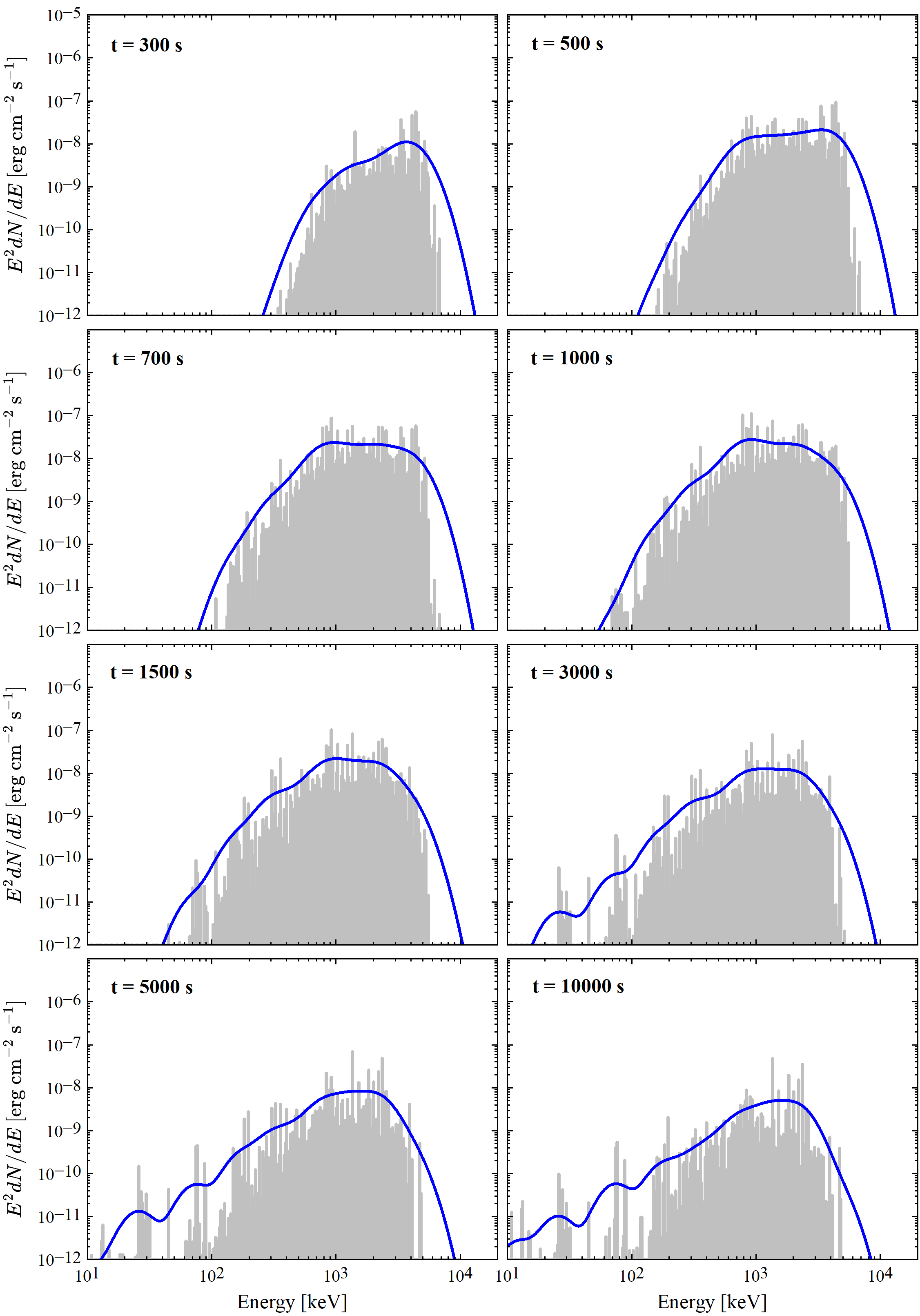}
    \caption{Spectrum of observed gamma-ray emission produced by radioactive nuclei in magnetar GFs. The gray vertical lines mark the rest-frame gamma-ray emission lines, while the blue curve shows the observed spectrum after accounting for Doppler broadening.}
    \label{spectral}
\end{figure*}

Based on our $r$-process nucleosynthesis simulations and the energy-dependent opacity, we compute the gamma-ray emission spectrum generated by radioactive nuclei in magnetar GFs, as shown in Figure~\ref{spectral}. Owing to the strong energy dependence of the gamma-ray opacity, the observed spectrum exhibits a significant delay at keV energies.
At $t = 300$~s, the gamma-ray emission is dominated by high-energy photons with $E_{\gamma} \gtrsim 1$~MeV.
Between $t = 300$ and $3000$~s, the MeV gamma-ray flux increases and reaches its peak, while the spectrum progressively extends toward the sub-MeV and keV bands. During the late phase ($t \sim 10^4$~s), the spectral evolution continues toward lower energies, accompanied by a steady increase in the low-energy flux. The evolution of these spectral features further highlights the crucial role of gamma-ray opacity. The time delay between MeV and keV photons suggests the presence of heavy nuclei in the ejecta.

\begin{figure*}
    \centering
    \includegraphics[width=0.75\textwidth]{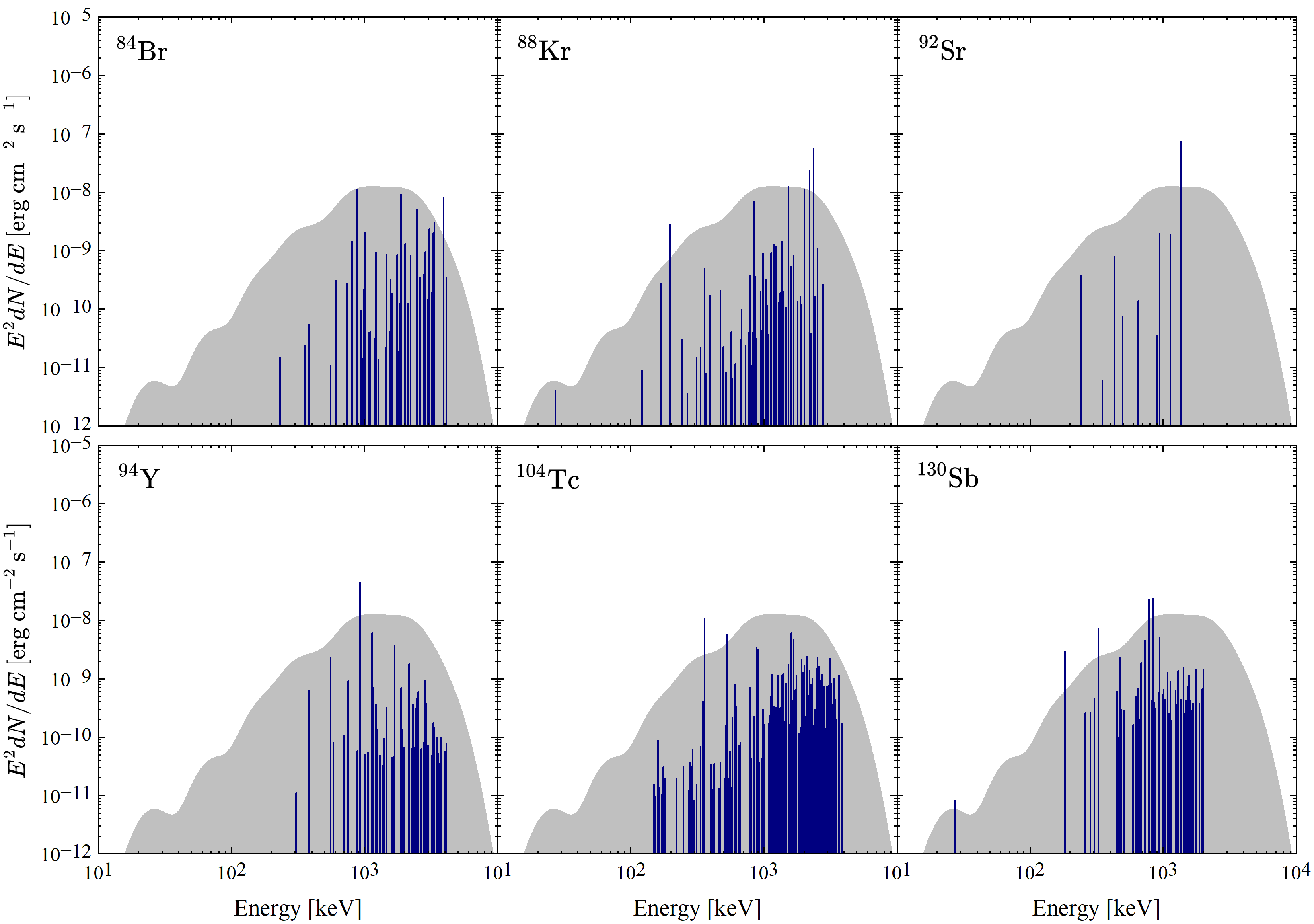}
    \caption{Gamma-ray emission lines powered by the dominant nuclides. The gray shaded region represents the observed spectrum from a magnetar GFs at $t = 3000$~s.}
    \label{dominant}
\end{figure*}

We further analyze the contributions of individual nuclides to the gamma-ray spectrum and identify the dominant contributors. By examining the energy generation rates of the radioactive decay chains, we find that nuclides near the first $r$-process peak dominate the gamma-ray emission, including $^{84}$Br, $^{87}$Kr, $^{88}$Kr, $^{88}$Rb, $^{89}$Rb, $^{92}$Sr, $^{94}$Y, $^{101}$Tc, and $^{104}$Tc. Heavy nuclei near the second $r$-process peak, such as $^{128}$Sn, $^{129}$Sb, and $^{130}$Sb, also contribute significantly. Figure~\ref{dominant} shows the gamma-ray emission lines produced by several of these dominant nuclides. The observed spectrum exhibits a prominent peak feature in the $\sim 1-2$~MeV energy range, primarily contributed by $^{88}$Kr, $^{92}$Sr, $^{94}$Y, and $^{130}$Sb. Among them, the radioacitve decay of $^{92}$Sr produces a bright gamma-ray emission line at $1.38$~MeV with a flux of $\sim 7 \times 10^{-8}$~erg~cm$^{-2}$~s$^{-1}$, while $^{88}$Kr produced two additional bright emission lines at $2.20$~MeV and $2.39$~MeV, each exceeding $2 \times 10^{-8}$~erg~cm$^{-2}$~s$^{-1}$. Notably, both $^{92}$Sr and $^{88}$Kr have half-lives of order $10^{4}$~s, allowing these bright gamma-ray lines to persist until $\sim 10^{4}$~s. Consequently, these nuclides are among the most promising candidates for direct identification in magnetar GFs with MeV gamma-ray detectors. 
The bright gamma-emission lines identified in this work are broadly consistent with those reported in previous work~\citep{2025ApJ...984L..29P}.
Table~\ref{line} lists the dominant nuclides responsible for bright gamma-ray emission lines along with their corresponding fluxes.

\begin{table*}[t!]
    \centering
    \caption{Dominant nuclides responsible for bright gamma-ray emission lines and their fluxes of the corresponding line.}
    \begin{tabular}{cccccc}
    \hline\hline
    Nuclide  &  Half-life~[s]  &  Parent nuclide  &  Parent half-life~[s]  &  Energy~[keV]  &  Flux~[erg~cm$^{-2}$~s$^{-1}$]\\
    \hline
    \multirow{3}{*}{$^{84}$Br}  &  \multirow{3}{*}{$1.91\times10^3$}  &  \multirow{3}{*}{$^{84}$Se}  &  \multirow{3}{*}{$1.96\times10^2$}  &  $881.60$  &  $1.11\times10^{-8}$  \\
      &  &  &  &  $1897.60$  &  $9.28\times10^{-9}$  \\
      &  &  &  &  $3927.50$  &  $8.28\times10^{-9}$  \\
    \hline
    \multirow{2}{*}{$^{87}$Kr}  &  \multirow{2}{*}{$4.58\times10^3$}  &  \multirow{2}{*}{$^{87}$Br}  &  \multirow{2}{*}{$55.65$}  &  $402.59$  &  $3.55\times10^{-9}$  \\
      &  &  &  &  $2554.75$  &  $6.36\times10^{-9}$  \\
    \hline
    \multirow{5}{*}{$^{88}$Kr}  &  \multirow{5}{*}{$1.02\times10^4$}  &  \multirow{5}{*}{$^{88}$Br}  &  \multirow{5}{*}{$16.29$}  &  $196.30$  &  $2.83\times10^{-9}$  \\
      &  &  &  &  $834.83$  &  $6.90\times10^{-9}$  \\
      &  &  &  &  $1529.77$  &  $1.28\times10^{-8}$  \\
      &  &  &  &  $2195.84$  &  $2.40\times10^{-8}$  \\
      &  &  &  &  $2392.11$  &  $5.49\times10^{-8}$  \\
    \hline
    \multirow{2}{*}{$^{88}$Rb}  &  \multirow{2}{*}{$1.07\times10^3$}  &  \multirow{2}{*}{$^{88}$Kr}  &  \multirow{2}{*}{$1.02\times10^4$}  &  $898.03$  &  $7.78\times10^{-9}$  \\
      &  &  &  &  $1836.00$  &  $2.51\times10^{-8}$  \\
    \hline
    \multirow{4}{*}{$^{89}$Rb}  &  \multirow{4}{*}{$9.09\times10^2$}  &  \multirow{4}{*}{$^{89}$Kr}  &  \multirow{4}{*}{$1.89\times10^2$}  &  $1031.92$  &  $1.61\times10^{-8}$  \\
      &  &  &  &  $1248.14$  &  $1.44\times10^{-8}$  \\
      &  &  &  &  $2195.92$  &  $8.00\times10^{-9}$  \\
      &  &  &  &  $2570.20$  &  $7.03\times10^{-9}$  \\
    \hline
    \multirow{1}{*}{$^{92}$Sr}  &  \multirow{1}{*}{$9.76\times10^3$}  &  \multirow{1}{*}{$^{92}$Rb}  &  \multirow{1}{*}{$4.49$}  &  $1383.93$  &  $7.44\times10^{-8}$  \\
    \hline
    \multirow{1}{*}{$^{94}$Y}  &  \multirow{1}{*}{$1.12\times10^3$}  &  \multirow{1}{*}{$^{94}$Sr}  &  \multirow{1}{*}{$75.30$}  &  $918.74$  &  $4.49\times10^{-8}$  \\
    \hline
    \multirow{1}{*}{$^{101}$Tc}  &  \multirow{1}{*}{$8.52\times10^2$}  &  \multirow{1}{*}{$^{101}$Mo}  &  \multirow{1}{*}{$8.77\times10^2$}  &  $306.83$  &  $4.83\times10^{-9}$  \\
    \hline
    \multirow{1}{*}{$^{104}$Tc}  &  \multirow{1}{*}{$1.10\times10^3$}  &  \multirow{1}{*}{$^{104}$Mo}  &  \multirow{1}{*}{$60.00$}  &  $358.00$  &  $1.07\times10^{-8}$  \\
    \hline
    \multirow{1}{*}{$^{107}$Rh}  &  \multirow{1}{*}{$1.30\times10^3$}  &  \multirow{1}{*}{$^{107}$Ru}  &  \multirow{1}{*}{$2.25\times10^2$}  &  $302.77$  &  $3.25\times10^{-9}$  \\
    \hline
    \multirow{5}{*}{$^{128}$Sn}  &  \multirow{5}{*}{$3.54\times10^3$}  &  \multirow{5}{*}{$^{128}$In}  &  \multirow{5}{*}{$0.84$}  &  $26.36$  &  $6.12\times10^{-11}$  \\
      &  &  &  &  $29.73$  &  $1.29\times10^{-11}$  \\
      &  &  &  &  $45.70$  &  $3.52\times10^{-11}$  \\
      &  &  &  &  $75.10$  &  $1.87\times10^{-10}$  \\
      &  &  &  &  $482.30$  &  $3.90\times10^{-9}$  \\
    \hline
    \multirow{4}{*}{$^{130}$Sb}  &  \multirow{4}{*}{$2.37\times10^3$}  &  \multirow{4}{*}{$^{130}$Sn}  &  \multirow{4}{*}{$2.23\times10^2$}  &  $182.3$  &  $2.93\times10^{-9}$  \\
      &  &  &  &  $330.9$  &  $7.07\times10^{-9}$  \\
      &  &  &  &  $793.4$  &  $2.29\times10^{-8}$  \\
      &  &  &  &  $839.5$  &  $2.40\times10^{-8}$  \\
    \hline
    \end{tabular}
    \label{line}
\end{table*}

\subsection{Implications for MeV Gamma-Ray Astronomy}

\begin{figure}[H]
    \centering
    \includegraphics[width=1.0\linewidth]{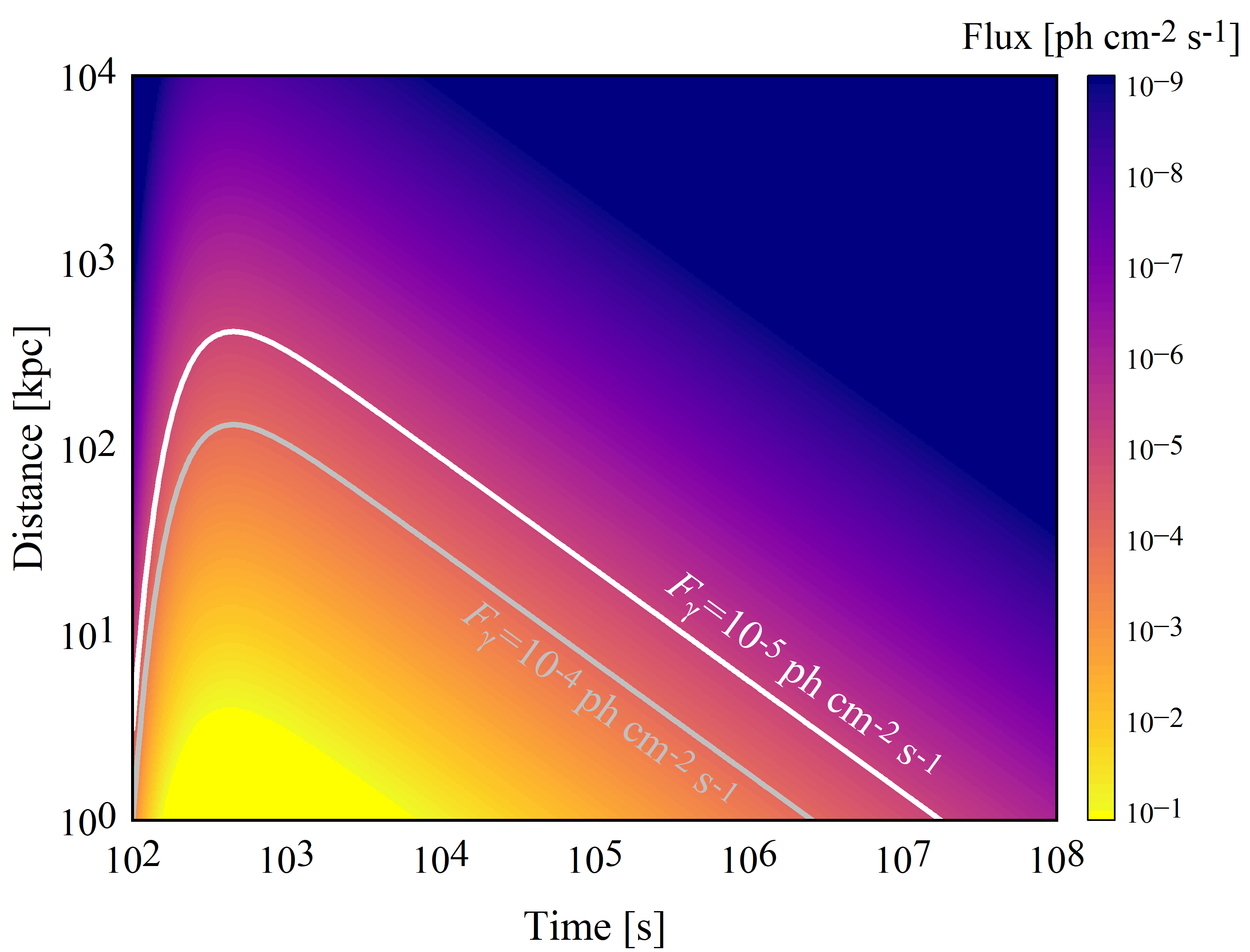}
    \caption{Detection prospects of nuclear gamma-ray lines from radioactive nuclei in magnetar GFs with MeV gamma-ray detectors. The gray and white curves indicate flux levels of $10^{-4}$ and $10^{-5}$~ph~cm$^{-2}$~s$^{-1}$, respectively, roughly corresponding to the sensitivities of current and next-generation MeV gamma-ray detectors.}
    \label{detection}
\end{figure}

Figure~\ref{detection} shows the expected detection prospects for nuclear gamma-ray lines arising from the radioactive decay of nuclei synthesized in magnetar GFs as observed by MeV gamma-ray detectors. We assume these flares occur at different distances and estimate the photon flux near 1~MeV at various times after the outburst. 
The sensitivities of current MeV gamma-ray instruments INTEGRAL~\citep{2003A&A...411L...1W} and next-generation missions such as the MeV Astrophysical Spectroscopic Surveyor (MASS)~\citep{2024ExA....57....2Z} are also indicated by the gray and white curves.
For a source at a given distance, the flux of nuclear gamma-ray lines initially rises and then declines, reaching its peak at $t \lesssim 10^{3}$~s. For magnetar GFs occurring within the Galaxy, the resulting gamma-ray emission is detectable by current MeV detectors with the detectable time window extending to $t\sim10^{5}$~s. In contrast, for flares occurring outside the Galaxy, detection of the associated nuclear gamma-ray lines is challenging and likely feasible only near the peak phase. 
For next-generation MeV gamma-ray instruments, including MASS~\citep{2024ExA....57....2Z}, COSI~\citep{2024icrc.confE.745T}, and GRAMS~\citep{2020APh...114..107A}, the line sensitivity could reach $\sim 10^{-5}$~ph~cm$^{-2}$~s$^{-1}$, making them highly promising for detecting gamma-ray lines from the radioactive decay of heavy elements and for identifying Galactic magnetar GFs.

\section{Summary and Discussion}
\label{Summary}

Recent studies have shown that magnetar giant flares (GFs) can eject $\sim 10^{-8}-10^{-6}~M_\odot$ of high-entropy and rapidly expanding baryonic material, thereby providing favorable conditions for $r$-process nucleosynthesis~\citep{2024MNRAS.528.5323C,2025ApJ...985..234P}. Subsequent modeling of nuclear gamma-ray emission arising from the radioactive decay of heavy nuclei in these flares has shown that the predicted late-time signatures are broadly consistent with the delayed gamma-ray features observed from the magnetar SGR 1806-20~\citep{2025ApJ...984L..29P}. These lines of evidence indicate that magnetar GFs can make a non-negligible contribution to the Galactic enrichment of heavy elements, and that the associated nuclear gamma-ray lines provide valuable targets for MeV gamma-ray astronomy.

In this work, we employ the nuclear reaction network code SkyNet to investigate the nucleosynthesis yields of magnetar GFs and to compute the gamma-ray opacity of the ejecta in detail based on the resulting nuclide composition. It is found that the nuclei synthesized in magnetar GFs are primarily distributed around the first and second $r$-process abundance peaks, corresponding to mass numbers in the range $80 \lesssim A \lesssim 130$. As a consequence of this nuclide composition, the gamma-ray opacity is found to be strongly energy-dependent, with the opacity in the keV band exceeding that in the MeV band by approximately three orders of magnitude. We note that the $r$-process nucleosynthesis results are obtained by summing over 30 layers with varying entropies and expansion timescales, so these parameters are expected to have only a minor effect on the total nucleosynthetic output. In addition, we set the initial electron fraction to $Y_{\rm e} = 0.46$, corresponding to that of iron-group nuclei, in agreement with the hydrodynamical simulations of magnetar GFs~\citep{2024MNRAS.528.5323C}. In cases of extremely strong magnetic fields, such as $B=10^{16.5}$~G, GFs can eject material from deeper layers of the neutron star crust, where the electron fraction can reach $Y_{\rm e} \sim 0.40$, providing more favorable conditions for the synthesis of heavy $r$-process elements~\citep{2024MNRAS.528.5323C}.

Based on the $r$-process nucleosynthesis simulations and the energy-dependent gamma-ray opacity, we investigate the nuclear gamma-ray emission from radioactive nuclei in magnetar GFs. The main features of the resulting gamma-ray emission can be summarized as follows:

\begin{itemize}

    \item The nuclear gamma-ray emission from magnetar GFs is dominated by MeV photons at early times ($t \lesssim 10^{3}$~s) and gradually extends toward the sub-MeV and keV bands as time progresses. This behavior arises because the photoelectric absorption of gamma-rays in the keV band is significantly enhanced in ejecta enriched with heavy elements, delaying their emergence relative to the MeV photons. Therefore, the time delay between the keV and MeV gamma-ray photons provides a diagnostic of heavy element enrichment in the ejecta.
    
    \item The gamma-ray spectrum of magnetar GFs exhibits a prominent feature near $\sim 1$~MeV with dominant contributions from $^{88}$Kr, $^{92}$Sr, $^{94}$Y, and $^{130}$Sb. In particular, the radioactive decay of $^{92}$Sr produces a bright gamma-ray line at 1.38~MeV with a flux of $\sim 7 \times 10^{-8}$~erg~cm$^{-2}$~s$^{-1}$, while $^{88}$Kr gives rise to two additional bright emission lines at 2.20~MeV and 2.39~MeV, each with fluxes exceeding $2 \times 10^{-8}$~erg~cm$^{-2}$~s$^{-1}$. We note that the half-lives of both $^{92}$Sr and $^{88}$Kr are of order $10^{4}$~s, allowing their bright gamma-ray lines to persist for up to $\sim 10^{4}$~s after the flare. The detection and identification of these nuclear gamma-ray lines would thus provide the most direct and robust observational evidence supporting magnetar GFs as astrophysical sites of $r$-process nucleosynthesis.
    
\end{itemize}

In our calculations of nuclear gamma-ray emission, we adopt an ejecta mass of $1.2 \times 10^{-6}~M_\odot$, consistent with the mass inferred from the gamma-ray emission of the magnetar SGR~1806-20~\citep{2025ApJ...984L..29P}. A simple Gaussian profile is used to account for Doppler broadening of the gamma-ray lines. While both the ejecta mass and the assumed line profile can influence the overall luminosity and detailed shape of the gamma-ray lines, they do not affect the main features of the predicted gamma-ray emission. This is because the characteristic spectral peaks and the time-delay effects of nuclear gamma-ray emission are primarily determined by the intrinsic properties of the ejecta composition rather than by the bulk ejecta parameters.

Detecting nuclear gamma-ray lines from radioactive nuclei in magnetar GFs requires a coordinated observational strategy combining wide–field gamma-ray monitors for triggering and localization with high–resolution gamma-ray spectrometers for follow-up observations and line identification. Currently operating wide–field instruments, such as the Space-based multiband astronomical Variable Objects Monitor (SVOM)~\citep{2016arXiv161006892W}, offer excellent capabilities for rapid follow-up and localization of MeV gamma-ray emission, providing early warning and precise positioning for magnetar GFs. Facilities with high line sensitivity like the International Gamma-Ray Astrophysics Laboratory (INTEGRAL)~\citep{2003A&A...411L...1W}, can then perform detailed follow-up observations and identify the nuclear gamma-ray lines, thereby diagnosing the elemental composition and temporal evolution of the ejecta. Furthermore, the proposed next-generation MeV gamma-ray spectrometers such as the MeV Astrophysical Spectroscopic Surveyor (MASS)~\citep{2024ExA....57....2Z}, are expected to achieve line sensitivities of $\sim 10^{-5}$~ph~cm$^{-2}$~s$^{-1}$ and energy resolutions of $\sim 0.6\%$ in the MeV band. Such advances would substantially enhance the prospects for the direct detection and unambiguous identification of nuclear gamma-ray lines, opening a new and unique observational window for probing $r$-process nucleosynthesis in magnetar GFs in the near future.

\vspace{2em}

\Acknowledgements{}
{{\noindent\bf Acknowledgements} This work is supported by the National Key R\&D Program of China (2024YFA1611704), the Guangxi Science and Technology Innovation Platform Program (Guike LT2600640026), and the National Natural Science Foundation of China (Grant Nos. 12403043, 12133003, 12473038, 12494574 and 12494570). MHC also acknowledges the support from the Guangxi Bagui Young Scholars Foundation. XGW acknowledges the support from the Guangxi Bagui Scholars Programme. This work is also supported by the Guangxi Talent Program (Highland of Innovation Talents).}

\vspace{1em}
\InterestConflict{}
{{\noindent\bf Conflict of interest} The authors declare that they have no conflict of interest.}



\bibliographystyle{apsrev4-1}
\bibliography{citation}

\end{multicols}

\end{document}